# Quantum transport in GaN/AlN double-barrier heterostructure nanowires

R. Songmuang[1], G. Katsaros[2], E. Monroy[1], P. Spathis[2], C. Bourgeral[1], M. Mongillo[2] and S. De Franceschi[2]

[1] CEA-CNRS group "Nanophysique et Semiconducteurs", Institut Néel/CNRS and CEA/INAC/SP2M, 17 rue des Martyrs 38054, Grenoble, France

[2] CEA, INAC/SPSMS/LaTEQS, 17 rue des Martyrs 38054, Grenoble, France

We investigate electronic transport in *n-i-n* GaN nanowires with and without AlN double barriers. The nanowires are grown by catalyst-free, plasma-assisted molecular beam epitaxy enabling abrupt GaN/AlN interfaces as well as longitudinal n-type doping modulation. At low temperature, transport in *n-i-n* GaN nanowires is dominated by the Coulomb blockade effect. Carriers are confined in the undoped middle region, forming single or multiple islands with a characteristic length of ~100 nm. The incorporation of two AlN tunnel barriers causes confinement to occur within the GaN well in between. In the case of 6-nm-thick wells and 2-nm-thick barriers, we observe characteristic signatures of Coulomb-blockaded transport in single quantum dots with discrete energy states. For narrower wells and barriers, Coulomb-blockade effects do not play a significant role while the onset of resonant tunneling via the confined quantum levels is accompanied by a negative differential resistance surviving up to ~150 K.

Corresponding author: rudeesun.songmuang@grenoble.cnrs.fr

silvano.defranceschi@cea.fr

Nitride (III-N) semiconductors are key materials for optoelectronic devices operating in the green to ultraviolet spectral region [1]. Their outstanding physical and chemical stability enables III-N devices to be operated under extreme environmental conditions, while their bio-compatibility and their piezoelectrical properties render them suitable for bio-chemical sensor applications. Furthermore, the large and adjustable band offsets of III-N heterostructures are a useful feature for devices relying on quantum confinement and tunneling transport. In addition, thanks to an efficient electron confinement, III-N quantum dots (QDs) are considered a promising system for the realization of single photon sources operating at room temperature [2].

The main problem faced by III-N semiconductors is that they are grown on non-native lattice-mismatched substrates, such as sapphire, SiC or Si. The structural defects resulting from plastic strain relaxation are known to deteriorate the performance of devices based on conventional two-dimensional heterostructures. Recently, III-N nanowires (NWs) have emerged as a promising route to circumvent this problem. The large surface-to-volume ratio of NWs favors an efficient elastic strain relaxation thereby inhibiting the generation of crystal misfit dislocations. At the same time, III-N NWs largely preserve the relevant properties of two-dimensional epitaxial growth, such as the possibility to form heterostructures [3] and to modulate both p- and n-type doping [4-5]. Because of these advantages, III-N NWs are being considered as a promising system for the development of new types of nano-scale photonic devices [6-8].

So far, experimental studies of III-N NWs have been focused mostly on their photonic properties [7-10], up to the recent demonstration of narrow excitonic emission from single quantum dots in GaN/AlN heterostructure NWs [10]. This Letter addresses the transport properties of n-type GaN/AlN double-barrier NWs. The realization of single quantum dot devices and resonant tunneling diodes (RTDs) out of these nanostructures is a first important step towards a variety of novel quantum devices relying on axial quantum confinement. To the best of our knowledge, the only experimental demonstration of a resonant-tunneling device made so far from heterostructure NWs was reported by Samuelson and collaborators using InAs/InP double-barrier NWs [11]. Of key importance in that work was the use of NWs with particularly abrupt InAs/InP heterojunctions, which were synthesized by catalytic chemical-beam epitaxy.

The III-N NWs used in this study were grown by plasma-assisted molecular beam epitaxy (PAMBE) on Si(111) without using any external catalysts [12-14]. This technique enables the growth of defect-free AlN/GaN NWs with abrupt heterojunctions [3,4]. These heterojunctions are obtained by switching the group-III element during the growth as opposed to the case of III-V heterostructure NWs realized by a catalytic method (e.g. InAs/InP NWs) where the group-V element is most efficiently switched.

A transmission electron micrograph (TEM) of a representative ensemble of as-grown NWs is shown in Fig. 1a. The NWs have diameters in the 50–80 nm range and typical length varying from sample to sample between 1.0 and 1.4 μm. Different NW samples were grown for this work, starting from a reference sample (A) made entirely of GaN with *n-i-n* doping profile (see top right panel of Fig. 1c); i.e. two Si-doped *n*-type edges and an undoped (*i*) middle section with a nominal length of 400 nm (this length is estimated from the vertical growth rate multiplied by the deposition time without taking Si dopant diffusion into account). The other samples consisted of double-barrier heterostructure NWs obtained by inserting a pair of thin AlN layers in the center of the undoped GaN region (see bottom right panel of Fig. 1c). The thickness of these layers and that of the GaN quantum dot in between were varied from sample to sample. Below we present data for two of such samples: sample B, with AlN barriers of 1.4±0.2 nm and a GaN quantum dot of 3.7±1 nm; sample C, with AlN barriers of 2.3±0.8 nm and a GaN quantum dot of 5.9±2 nm. These thicknesses were determined from high-resolution TEM (HRTEM) over an ensemble of 5-6 wires for each sample. Figure 1b shows a representative HRTEM image of a GaN/AlN double-barrier heterostructure NW from sample C. The abruptness of the GaN/AlN interfaces is apparent. Also a small lateral growth of a few-nm-thick GaN shell around the AlN barriers can be seen as indicated by the arrows. Due to surface depletion, lateral conduction through this shell was found to be negligible as discussed below.

In order to perform electrical transport measurements, the NWs were mechanically dispersed on heavily doped Si substrates capped with a 285-nm-thick $SiO_2$ layer [Left panel of Fig. 1(c)]. Some selected NWs were individually contacted with Ti/Al/Ti/Au electrodes defined by e-beam lithography. To make sure that the undoped NW section is comprised between the source and drain contacts, the distance between them was chosen to be in the 400–600 nm range. Transport measurements were performed by means of a standard lock-in technique with the sample mounted in a $He^4$ cryostat with adjustable temperature between 4.2 and 300 K.

Figure 2a shows the source-drain current ($I_{SD}$) as a function of the applied bias voltage ($V_{SD}$) for a device fabricated from a reference *n-i-n* GaN NW (sample A). A linear characteristic is observed at room temperature (red trace). This linear behavior has been reproduced in 26 devices with typical resistances of a few MΩ. (Similar test devices fabricated from uniformly doped *n*-type GaN NWs exhibited resistances an order of magnitude lower.)

The intentional lack of extrinsic doping in the middle region of the NW enables the device resistance to be varied by a gate voltage ($V_G$) applied to the doped-Si substrate. Figure 2b shows an example of two $I_{SD}(V_G)$ characteristics obtained on the same *n-i-n* device for both sweeping directions of $V_G$. The device is completely turned off for sufficiently negative $V_G$, confirming the n-type character of the NW. The threshold gate voltage is found to depend on the sweeping direction. This hysteretic behavior, frequently observed in NW field-effect devices, is likely due to charge traps on the NW surface (the shown measurements were taken in vacuum).

From an average threshold voltage $V_{G,th}$ = –5 V, we can roughly estimate the carrier concentration ($N_e$) in the undoped region of the NW at $V_G$ = 0. A simplified electrostatic model [15] yields $N_e = 2\varepsilon\varepsilon_0 V_{G,th}/er^2\ln(2h/r)$ ~1x10$^{18}$ cm$^{-3}$, where $e$ is the electron charge, $\varepsilon_0$ the vacuum permittivity, $r$ the NW radius; $\varepsilon$ and $h$ are the dielectric constant and the thickness of the substrate oxide, respectively. For most of the devices, however, the NW channel could not be completely depleted (i.e. $V_{G,th}$ < –30 V), denoting carrier concentrations of ~10$^{19}$ cm$^{-3}$, i.e. largely exceeding the background doping level (10$^{17}$ cm$^{-3}$) of undoped GaN films grown in the same machine. This large discrepancy may be due to residual *n*-type doping coming from the incorporation of O impurities, enhanced by the N-rich growth conditions, and to the diffusion of Si atoms along the NW axis, favored by the high growth temperature [16].

At 4.2 K, the $I_{SD}(V_{SD})$ characteristics of *n-i-n* GaN NWs develop a non-linear behavior as shown by the blue trace in Fig. 2a. The observed suppression of $I_{SD}$ around zero bias is a signature of the Coulomb blockade effect which becomes dominant at low temperature. A clear understanding of this regime is obtained from a measurement of the differential conductance ($dI_{SD}/dV_{SD}$) as a function of ($V_G, V_{SD}$). As shown in Fig. 2c, the Coulomb blockade effect results in a sequence of diamond-shape dark regions in which transport is entirely suppressed. This behavior, characteristic of single-electron tunneling, implies that the undoped region of the NW confines a single electronic island. Within each Coulomb diamond, the island contains a well-defined, integer number of electrons. The charging energy $E_C$ associated with the addition of an electron to the island is around a few meV, as it can be estimated from the height of the Coulomb diamonds. The diamond size $\Delta V_G$, measured along the horizontal axis, gives a coupling capacitance to the back gate, $C_G = e/\Delta V_G \sim 3$ aF, which corresponds to an island length of ~100 nm, i.e. smaller but comparable to the extension of the undoped region. This denotes a relatively low amount of crystal disorder. In fact, disorder can lead to the formation of multiple electronic islands in series with each other. Multiple-island configurations give rise to highly irregular and overlapping Coulomb diamonds, as indeed observed in other gate-voltage ranges and in different devices.

At least two different strategies exist to overcome the effect of disorder and obtain quantum dots with well-defined tunnel barriers, namely local charge depletion by means of narrow gates crossing the NW [17] and incorporation of thin barrier layers made of a larger gap semiconductor [12,18-21]. Here we focus on the latter approach investigating the formation of quantum dots in GaN by means of two closely spaced AlN layers. The first step is to show that AlN can form an efficient tunnel barrier and that the side-wall growth of GaN does not create a leakage path. To this aim, a test sample was grown consisting of *n-i-n* GaN NWs incorporating a single 10-nm-thick AlN layer. This layer was found to be thick enough to prevent electron tunneling. At the same time, the total absence of conduction proved that the laterally grown GaN shell is completely depleted and it does not lead to any measurable leakage.

Following this preliminary test, we considered devices with built-in AlN double barriers. Figure 3a shows a comparison of the low-temperature $I_{SD}(V_{SD})$ characteristics of a single *n-i-n* GaN NW (sample A) and those of two GaN NWs from sample B having closely-spaced AlN double barriers. The insertion of these barriers results in a much stronger nonlinearity, with $I_{SD}$ being asymmetrically suppressed over a larger bias-voltage scale. For example, in device 1, $I_{SD}$ is completely suppressed for $V_{SD}$ between -0.08 and 0.05 V. In addition, negative differential resistance (NDR) features clearly appear at both negative and positive bias, as indicated by the arrows. We ascribe these features to resonant tunneling through the discrete levels resulting from axial quantum confinement between the AlN barriers.

To estimate the energies of these levels we have used a one dimensional self-consistent Poisson-Schrödinger solver using Nextnano[3] simulation package [22]. This approximation is justified by the relatively small effect of lateral confinement, which is expected from the relatively large NW diameter as compared to the GaN quantum-dot height.

Figure 3b shows the calculated one-dimensional energy diagram of a double-barrier heterostructure with 1.5-nm-thick AlN barriers and a 4-nm-thick GaN dot along the [0001] direction (NW growth direction). Based on our previous field-effect measurements on *n-i-n* GaN NWs, a uniform doping density of $7 \times 10^{18}$ cm$^{-3}$ was assumed in this calculation. In addition, the full elastic strain relaxation of AlN/GaN heterostructures was used. As a consequence, the band bending profile is caused only by the spontaneous polarization difference between wurtzite AlN and GaN.

Size quantization along the axial direction yields a set of seven discrete levels. Only the first three levels, labeled as $E_1$, $E_2$, and $E_3$, are shown in Fig 3b. These levels lie at 0.20, 0.48, and 0.78 eV above the Fermi energy ($E_F$), respectively. They define the minima of three two-dimensional subbands resulting from the assumed free motion of electrons in the transverse plane.

We attribute the observed NDR features to the onset of tunneling via the different subbands. As in conventional resonant-tunneling diodes, NDR occurs when a quantized level lines up with the conduction band of the injecting contact [23]. The relatively large quantization energies resulting from such a small quantum dot, height explain why the NDR features are observed at bias voltages on the scale of hundreds of mV. Yet the model used does not allow us to establish an unambiguous correspondence between the NDR features and the calculated quantized levels. The effect of surface states and the possible presence of the remained strain resulting from the lattice mismatch between GaN and AlN are not taken into account. Moreover, significant uncertainties exist over critical parameters such as the quantum-dot height and the doping concentration which can vary from wire to wire. In fact, the presence of a small but measurable conduction at bias voltages smaller than 0.2 V suggests that $E_1$ may lie closer to the Fermi energy than in Fig. 3b.

NDR features are reproducibly observed both for both sweeping directions of $V_{SD}$. The most pronounced NDRs have peak-to-valley current ratios of 1.5 at 4.2 K and they can be resolved up to 150 K. No degradation of the device characteristic is found after repeated measurements, in contrast to what has been reported for AlN/GaN RTDs based on conventional two dimensional layers [24-26].

The pronounced asymmetry in the $I_{SD}(V_{SD})$ characteristics, shown in Fig. 3a and consistently observed in all the devices measured, can be ascribed to the difference polarization field at GaN/AlN interface causing the depletion and the electron accumulation region at the two sides of the AlN double barrier. We argue that the best matching between the data in Fig. 3a and the indicative calculation in Fig. 3b is obtained by assuming that the bias voltage is applied to the right edge of the NW (where a depletion region forms). In this case a positive bias increases the depletion region adjacent to the right-most heterojunction. Following the above conjecture that $E_1$ is close to $E_F$, the two clearly visible NDR features would correspond to the onset of resonant tunneling through $E_2$ and $E_3$. For negative bias voltages, only the NDR associated with $E_3$ would be visible since $E_2$ lies well below the conduction band edge on the injecting contact. This description is further supported by the steep rise of the current on the negative-bias side which is consistent with the disappearance of the depletion region on the right (i.e. the applied voltage falls entirely on the double-barrier).

The $V_{SD}$ position of the NDR features can be tuned by the back-gate voltage. This effect, shown if Fig. 3c, is the result of an energy shift of the resonance energy level induced by a variation of the local electrostatic potential. Accordingly, the NDR $V_{SD}$ position shifts to less negative bias voltages when $V_G$ is increased. Due to the onset of leakage towards the back gate, for $V_G$ exceeding ~35 V, the NDR could not be shifted all the way to zero bias.

The essential requirement for a sizeable conduction at small bias voltages is to have quantized energy levels at or below $E_F$. In GaAs-based resonant-tunneling devices developed by Tarucha and collaborators this condition was realized by adding a small amount of In the GaAs quantum well, since the smaller gap of InGaAs compensates for the positive energy shift due to quantum confinement [20]. In the III-N nanostructures considered here, the triangular shape of the conduction band profile along the NW growth axis resulting from the spontaneous polarization fields enables the energy of the lowest level $E_1$ to be lowered below $E_F$ simply by increasing the quantum-dot height. This principle is illustrated in Fig. 4a, where the energy of the first three levels is plotted as a function of the GaN quantum-dot height. Note that we do not take the strain induced electric field in to account for this calculation.

Following this principle we considered similar double-barrier NWs with a thicker GaN quantum dot (~6 nm) and slightly thicker AlN barriers (sample C). Figure 4b shows two $I_{SD}(V_{SD})$ characteristics at 4.2 K and different gate voltages for a device fabricated from one of these NWs The gate dependent suppression of $I_{SD}$ on a 10-mV range around zero bias is due to the Coulomb blockade effect. This is confirmed by the conductance oscillations in Fig. 4c (visible up to 35 K), and by the characteristic diamond-shape features in the color plot of $dI_{SD}/dV_{SD}(V_G,V_{SD})$ shown in of Fig. 4d.

The asymmetric shape of the Coulomb diamonds denotes a different capacitive coupling of the island to the source and the drain leads. This can be ascribed to the asymmetric conduction-band profile induced by the polarization fields. The characteristic size of the Coulomb diamonds gives a charging energy of ~10 meV and a gate capacitance $C_G \sim 0.1$ aF, corresponding to an island size of a few nm, i.e. much smaller than the one found for *n-i-n* GaN NWs. This size matches reasonably well with the height of GaN quantum dot. This correspondence is confirmed by other measurements performed on similar double-barrier devices which gave consistent values of $C_G$. We thus conclude that the observed single-electron transport occurs through an electronic island defined by the two AlN barriers.

A closer look at the data of Fig. 4d reveals the presence of additional structures appearing as lines (i.e. $dI_{SD}/dV_{SD}$ peaks) parallel to the diamond edges. These lines can be ascribed to the onset of single-electron tunneling through some excited states of the electronic island in the GaN quantum dot. Their presence constitutes direct evidence of a discrete energy spectrum: the motion of electrons within the well is quantized not only along the NW axis but also in the transverse plane due to the finite NW diameter. From the separation between the observed excited-state lines and the corresponding diamond edges, we estimate the level spacings to be in the range of 1–10 meV. Because the NW diameter is an order of magnitude larger than the GaN quantum dot height, these energy spacings are determined by the size quantization in the transverse plane. A simple estimate of the expected mean level spacing gives $\Delta E \sim 2\hbar^2/m^*r^2 \sim 1$ meV, consistent with the experimental finding. In this estimate we have used $m^* = 0.2 m_e$ for the effective mass of conduction electrons in wurtzite GaN ($m_e$ is the free electron mass), and $r = 25$ nm for the NW radius. Note that the effective radius may be somewhat smaller than the NW radius due to surface depletion.

To the best of our knowledge there have been only a few reports on quantum-dot electronic devices based on GaN. In those earlier studies the GaN quantum dots were created either by top gating a two-dimensional electron gas buried in a GaN/AlGaN heterostructure [27-28], or by fabricating nano-gap contacts to Stranski-Krastanow self-assembled GaN quantum dot [29].

The single-electron tunneling regime shown in Fig. 4d is entirely washed out at a higher $V_{SD}$ scale due to an increase in the transparency of the tunnel barriers and to the larger phase space for inelastic tunneling processes. NDR features associated with the onset of resonant tunneling through higher-energy subbands have been only occasionally observed.

In conclusion, we have presented the first electronic transport study of GaN/AlN axial-heterostructure NWs. We have shown that quantum dots can be defined in a GaN NW by incorporating two thin and closely-spaced AlN layers acting as tunnel barriers. Different transport regimes are observed at low temperature depending on the thicknesses of quantum-dot and the barriers. NDR features due to resonant tunneling, which are observed at relatively large bias voltages (0.1 – 1 V), provide experimental evidence of longitudinal energy quantization. Instead, transverse energy quantization is revealed by single-electron tunneling transport at a lower bias-voltage scale (0 – 20 mV). Our results show that axial-heterostructure III-N NWs are a viable and versatile option for exploring new nano-scale quantum devices.

The authors acknowledge H. Mariette for helpful discussions and the PTA cleanroom team for support in device fabrication. This research was partly funded by the Agence Nationale de la Recherche through the COHESION project. G. K. acknowledges further support from the Deutsche Forschungsgemeinschaft (grant KA 2922/1-1).

**Figure captions**

Figure 1. (a) TEM image of GaN NWs grown on a Si(111) substrate. (b) HRTEM image a NW section containing a double-barrier heterostructure, i.e. a pair 2-nm-thick AlN barriers (bright contrast) separated by 6-nm-height GaN quantum dot. The GaN lateral growth around the AlN barriers is clearly seen indicated by the arrows. (c) Schematic illustrations of a single NW based device with metal contacts on an oxidized Si substrate (left). The top (bottom) right scheme shows an *n-i-n* NW without (with) AlN double tunnel barriers.

Figure 2(a) Current-voltage characteristic of a single *n-i-n* NW at room temperature (red trace) and at 4.2K (blue trace). (b) $I_{SD}$ plotted as a function of $V_G$ for an *n-i-n* GaN NW with $V_{SD}$=10 mV when the $V_G$ was swept forward (red trace) and backward (blue trace). (c) Color scale plot of the differential conductance $dI_{SD}/dV_{SD}$ versus ($V_G$, $V_{SD}$). The measurement was taken at 4.2K.

Figure 3(a) $I_{SD}$-$V_{SD}$ characteristics at 4.2K of GaN NWs with two closely spaced AlN tunnel barriers in comparison to that of *n-i-n* GaN NWs. (b) a 1D conduction band diagram of a 1.5nm AlN / 4 nm GaN / 1.5nm AlN structure (c) $I_{SD}$-$V_{SD}$ characteristic showing the evolution of the NDR appearing at negative $V_{SD}$ for different $V_G$.

Figure 4 (a) Plot showing the one dimensional calculation of three lowest quantized levels for a GaN quantum dot between two 2-nm thick AlN barriers as a function of the dot height. (b) $I_{SD}$-$V_{SD}$ characteristic of a GaN NW with two AlN tunnel barriers (sample C) at different $V_G$ (c) Differential conductance versus $V_G$, revealing coulomb blockade peaks. The measurement was done by using the lock-in technique with an alternate frequency of 13.305 Hz and an excitation amplitude of 500µV. (d) Color scale plot of $dI_{SD}/dV_{SD}$ versus $V_G$ and $V_{SD}$. All measurements were taken at 4.2K

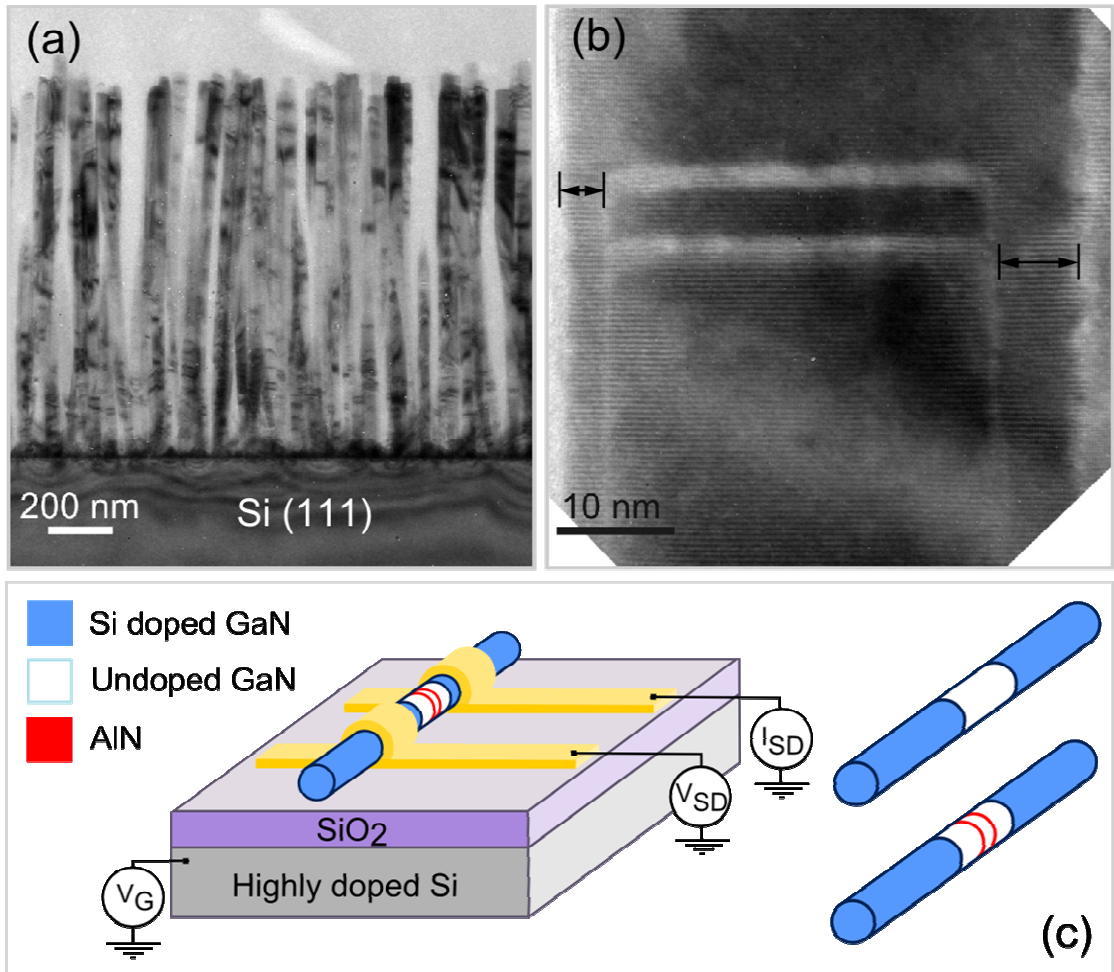

Figure 1

R. Songmuang *et al*.

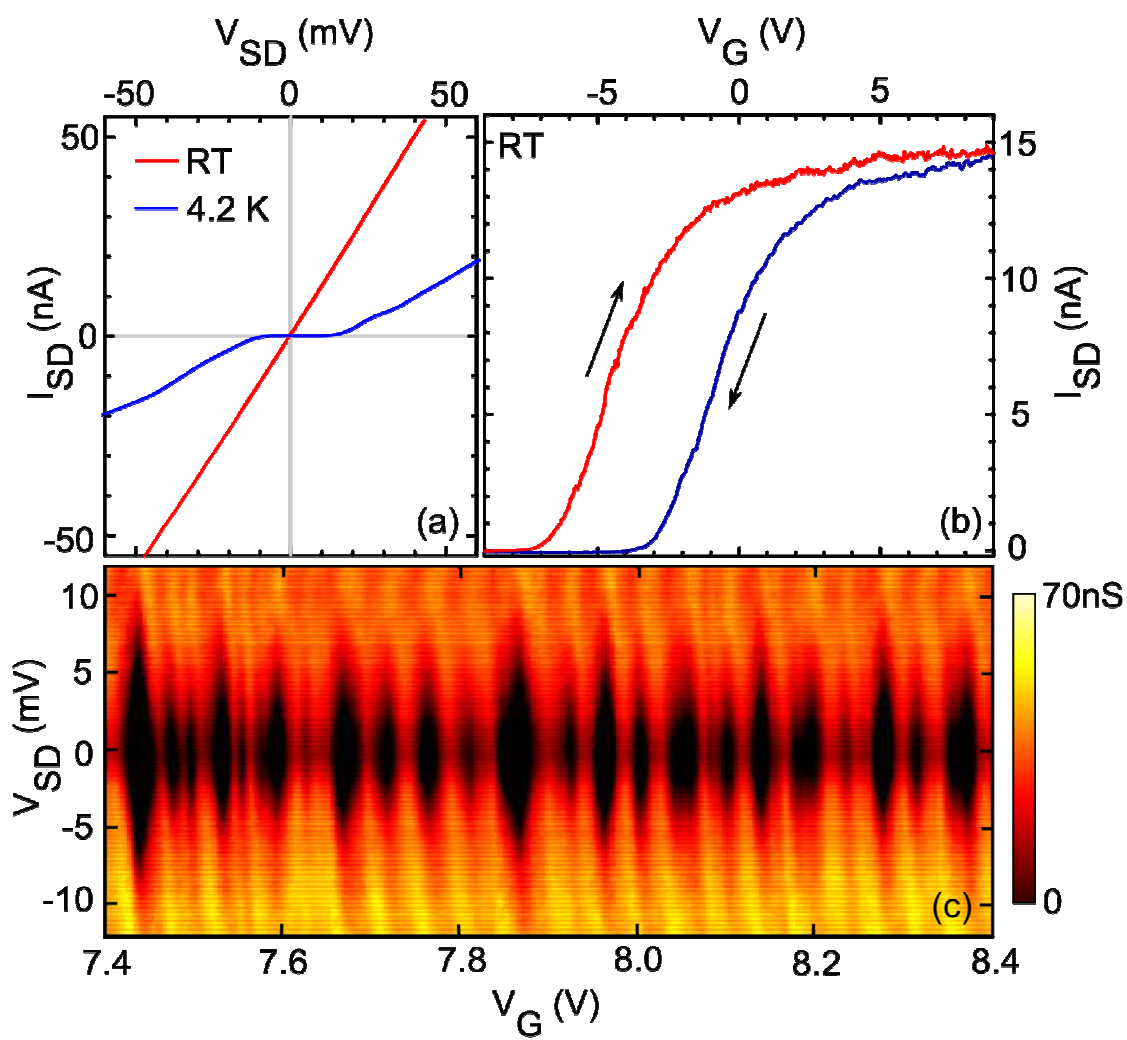

Figure 2

R. Songmuang *et al*.

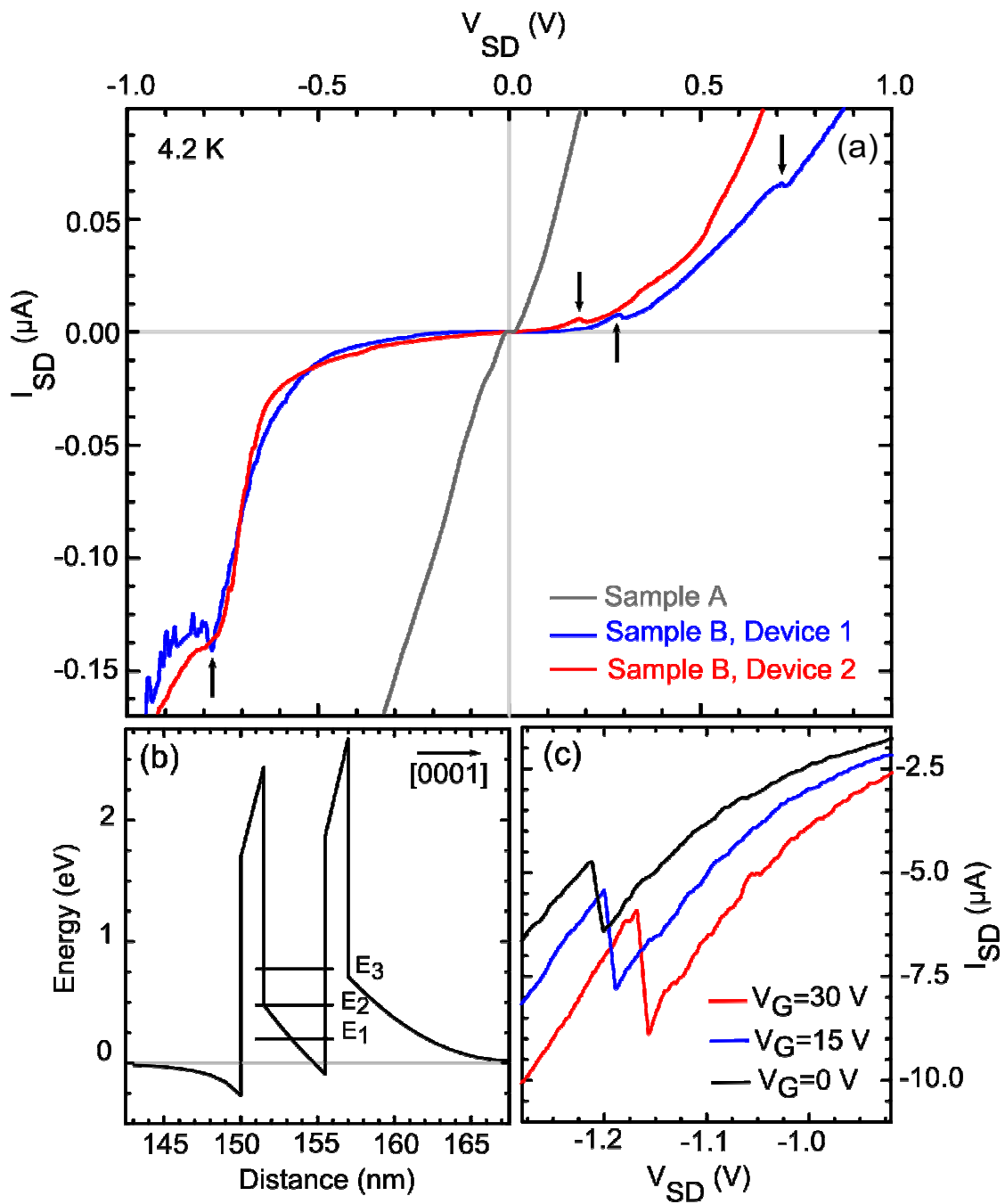

Figure 3

R. Songmuang *et al*.

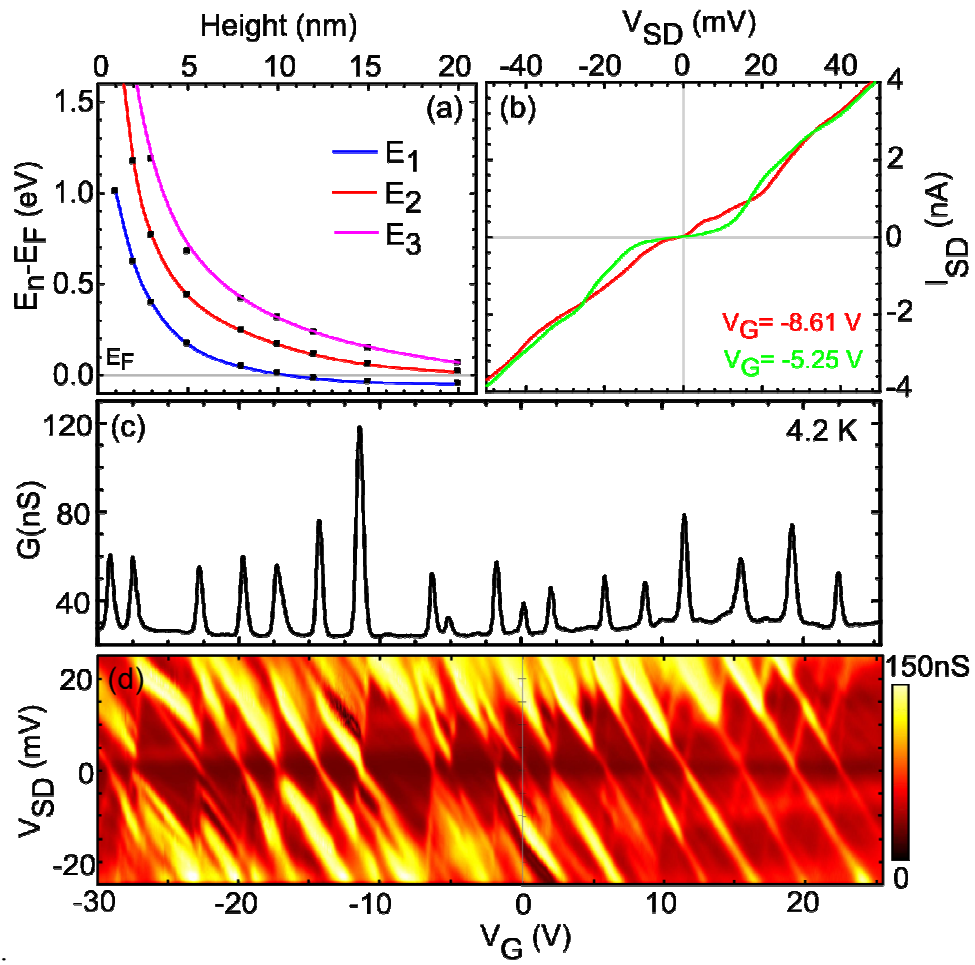

Figure 4

R. Songmuang *et al*.